\input{epsf.sty}
\documentstyle[12pt]{article}

\def\dofig#1{\vskip.2in\centerline{\epsfbox{#1}}}

\def\simge{\mathrel{%
   \rlap{\raise 0.511ex \hbox{$>$}}{\lower 0.511ex \hbox{$\sim$}}}}
\def\simle{\mathrel{
   \rlap{\raise 0.511ex \hbox{$<$}}{\lower 0.511ex \hbox{$\sim$}}}}
 
\def\slashchar#1{\setbox0=\hbox{$#1$}           % set a box for #1
   \dimen0=\wd0                                 % and get its size
   \setbox1=\hbox{/} \dimen1=\wd1               % get size of /
   \ifdim\dimen0>\dimen1                        % #1 is bigger
      \rlap{\hbox to \dimen0{\hfil/\hfil}}      % so center / in box
      #1                                        % and print #1
   \else                                        % / is bigger
      \rlap{\hbox to \dimen1{\hfil$#1$\hfil}}   % so center #1
      /                                         % and print /
   \fi}                                         %
\def\ts{\thinspace}

\def\ra{\rightarrow}

\def\ol{\bar}

\def\be{\begin{equation}} 
\def\ee{\end{equation}} 
\def\bea{\begin{eqnarray}}
\def\eea{\end{eqnarray}}
\def\ba{\begin{array}}
\def\ea{\end{array}}

\def\etmiss{\slashchar{E}_T}

\def\ecm{\sqrt{s}}

\def\Ntc{N_{TC}}

\def\kslash{\raise.15ex\hbox{/}\kern-.57em k}
\def\tro{\rho_{T}}

\def\tropm{\rho_{T}^\pm}

\def\troz{\rho_{T}^0}
\def\tpi{\pi_T}
\def\tpipm{\pi_T^\pm}

\def\tpiz{\pi_T^0}

\def\dfjj{\Delta\phi(jj)}
\def\wjj{Wjj}

\def\jet{\rm jet}
\def\jets{\rm jets}

\def\gev{{\rm GeV}}
\def\tev{{\rm TeV}}

\def\pb{{\rm pb}}

\def\ifb{{\rm fb}^{-1}}

\begin{document}
\title{
\vskip -15mm
\begin{flushright}
\vskip -15mm
{\small 
FERMILAB-PUB-99/002\\
hep-ph/9901202\\}
\vskip 5mm
\end{flushright}
{\Large{\bf \hskip 0.38truein
Can a light technipion be discovered at the Tevatron
if it decays to two gluons?}}\\
}
\author{
{\small Stephen Mrenna$^{1}$\thanks{mrenna@physics.ucdavis.edu}, and}
{\small John Womersley$^{2}$\thanks{womersley@fnal.gov}}\\
{\small {$^{1}$}University of California, Davis, CA 95616}\\
{\small {$^{2}$}Fermi National Accelerator Laboratory, Batavia, IL 60510}\\
}
\maketitle
%%%%%%%%%%%%%%%%%%%%%%%%%%%%%%%%%
%\vspace*{.3in}
\begin{abstract}
In multiscale and topcolor--assisted models of walking technicolor,
light, spin--one technihadrons can exist with masses of a few
hundred GeV; they are
expected to decay as $\tro \ra W\tpi$.
For $M_{\tro} \simeq 200\,\gev$ and $M_{\tpi} \simeq 100\,\gev$, the
process $p\bar p\to \tro\to W\tpi$
has a cross section of about a picobarn at the Tevatron. 
We demonstrate the detectability of this process
with simulations appropriate to Run~II
conditions, for the challenging case where 
the technipion decays dominantly into two gluons.
\end{abstract}

%%%%%%%%%%%%%%%%%%%%%%%%%%%%%%%%%
%%%%%%%%%%%%%%%%%%%%%%%%%%%%%%%%%

\newpage
\parskip \baselineskip

Color--singlet technipions, $\tpipm$ and $\tpiz$, are the
pseudo--Goldstone bosons of multiscale technicolor \cite{multi} and
topcolor--assisted 
technicolor~\cite{topcondref,topcref,tctwohill,tctwoklee}, and are
expected to be the lightest particles associated with the new physics.
These technipions couple to fermions in a similar fashion as
the standard model Higgs boson, with a magnitude set by
the technipion
decay constant $F_T$, but have no tree level couplings
to $W$ or $Z$ gauge bosons.   
Consequently, technipions can be produced singly at
hadron colliders through tree--level quark annihilation or 
one--loop gluon fusion processes.
If the quark annihilation process is dominant, then the production
rate is feeble, and the signal is a dijet peak, maybe with heavy flavor.
There is little hope of distinguishing this from QCD heavy flavor backgrounds.
If gluon fusion is dominant, then the production rate can be large,
but the signal is a digluon peak, with daunting backgrounds, or 
a rare $\gamma\gamma$ peak.  It may be possible to observe the latter
at the Large Hadron Collider (LHC) or possibly even the Tevatron, but
this is still an open question.

On the other hand, technivector mesons also arise in technicolor models,
and these can be produced at substantial rates through their mixing
with gauge bosons~\cite{tpitev,snow}. The technivector mesons in question are an
isotriplet of color-singlet $\tro$ and the isoscalar partner $\omega_T$.
Because techni-isospin is likely to be a good approximate symmetry, $\tro$
and $\omega_T$ should have equal masses.
% as do the various technipions. 
The enhancement of technipion masses due to walking technicolor~\cite{wtc}
suggests that the channels $\tro \ra \tpi\tpi$ and $\omega_T \ra
\tpi\tpi\tpi$ are kinematically closed. Thus, the decay modes $\tro \ra W_L \tpi$ and
$Z_L \tpi$, where $W_L$, $Z_L$ are longitudinal weak bosons, and $\omega_T
\ra \gamma \tpi$ may dominate~\cite{multi}.  In recent phenomenological 
analyses, it has been assumed that
$\tpiz$ decays into $b \ol b$. The presence of heavy flavor, plus an isolated
lepton or photon, provides clear signatures for these processes.
It has been demonstrated that such signals can be easily
detected in Run~II of the Fermilab Tevatron~\cite{elw}, and experimental 
searches following this prescription have been carried out on 
the Run~I dataset~\cite{cdfsearch}. 

However, it is quite possible that light technipions contain colored
technifermions, in which case
the decay $\tpiz \to gg$ may contribute significantly or dominate if
the number of technicolors $\Ntc$ is large \cite{colorref,gunion}.  
In this case,
the signature of technivector production is $\gamma$ or $W+2$~jets, 
and the backgrounds are correspondingly more severe.   
We present simulations of $\ol p p \ra \tropm \ra W^\pm_L \tpiz$
for the Tevatron collider with $\sqrt{s} =
2\,\tev$ and an integrated luminosity of $2\,\ifb$, corresponding to
that expected in Run~II.  We follow the previous analysis which
studied $\pi^0_T\to b\bar b$~\cite{elw} in using
topological cuts to exploit the resonant production process and thus
enhance the signal--to--background ratio. 
For the cross sections and cuts we use in this paper,
the signal stands out well above the background.
We would expect that the complementary channel
$\ol p p \ra \troz \ra Z_L \tpiz$ with $Z \to \ell^+ \ell^-$ might add
some sensitivity, but we have not considered this in detail. For
the case of $Z \to \nu \bar\nu$, the fake $\etmiss$ background from 
QCD multijet production is likely to
overwhelm the signal.

We have used {\sc Pythia}~6.1~\cite{pythia} to
generate $\ol p p \ra \tropm \ra W^\pm \tpiz$ and $\tpiz \ra gg$ 
at the Tevatron Collider with $\ecm = 2\,\tev$. 
The cross section is calculated
under the assumptions of~\cite{elw}.
The total production cross section times the branching ratio for
the decay of the $W$--boson into electrons, $\sigma\cdot B(W^\pm \to e^\pm)$,
is in the range 0.1--0.7~pb for $m_{\tro} \simge 200$~GeV and
$m_{\tpi} \simge 100$~GeV.  The parameter dependence is illustrated
in Fig.~\ref{plot0}.  
For detailed studies we have focused on a typical point, with
$m_{\tro} =
220\,\gev$ and $m_{\tpi} = 110\,\gev$, where 
$\sigma\cdot B = 0.45\,\pb$. 

\begin{figure}
\epsfxsize=15cm
\dofig{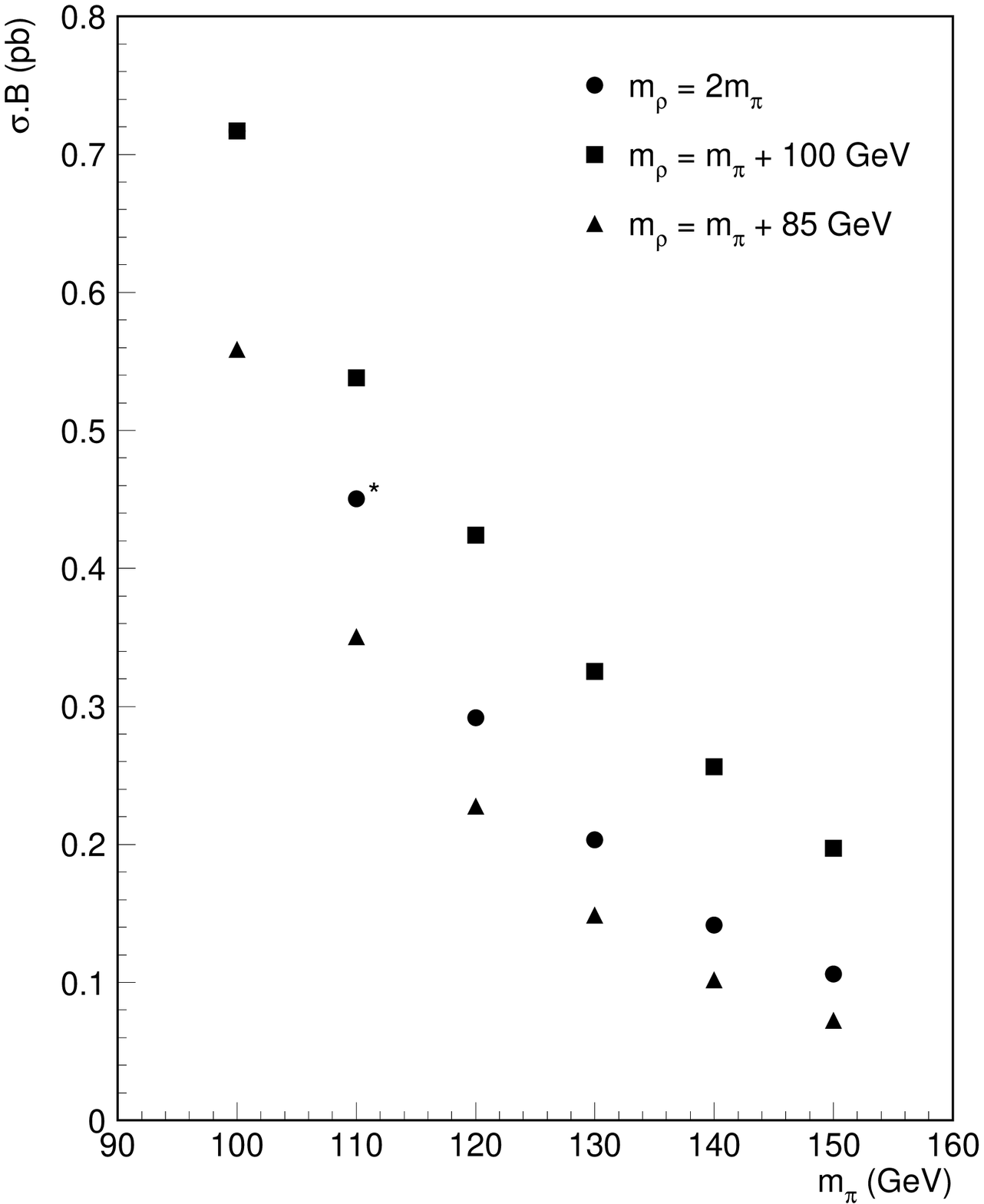}
\caption[]
{Production cross section for $p \overline p \to \rho_T \to \tpiz W$,
multiplied by $B(W\to e \nu)=0.105$, for various masses of $\tro$ and
$\tpi$.  The asterisk denotes the point ($m_{\tro} = 220, m_{\tpi}=110$~GeV)
which was studied in detail.
\label{plot0}}
\end{figure}

The partial width for the decay $\pi^0_T\to gg$ can be
calculated using the (energy--dependent) 
expression 
\begin{eqnarray}
\Gamma(\pi_T\to gg) = {1\over 128 \pi^3 F_T^2}\alpha_s^2 C_{\pi_T} N^2_{TC} s^{3/2},
\end{eqnarray}
while the competing decay has the width
\begin{eqnarray}
\Gamma(\pi_T\to b\bar b) = {1\over 4 \pi^2 F_T^2} 3 p_f C_f m_b^2.
\end{eqnarray}
In these expression, $\sqrt{s}$ is the resonance mass, $p_f$ is the $b$--quark
momentum in the resonance rest frame, and the constants $C_{\pi_T}$  and $C_f$ 
account for the flavor content of the technipion wave function.  By comparing
these expressions, we find that the $\pi_T^0\to gg$ decay begins to dominate
when $\Ntc\simge 3-4$.
For the purposes of this study, we have forced the $\pi^0_T$ to decay always into
the $gg$ final state.  Our final results refer to the quantity $\sigma\cdot B$ denoted
above times the branching ratio for $\pi^0_T\to gg$.

The dominant background, $W^\pm \ts\jet\ts\jet$, was calculated 
in two ways: firstly using the standard implementation of
{\sc Pythia} for the processes $qg \to Wq^\prime$ and 
$q\overline q^\prime \to Wg$ with a minimum $W$ transverse momentum of
15~GeV/$c$; and secondly by interfacing the explicit ``2 to 3'' processes
({\it e.g.} $q\overline q^\prime \to Wgg$, $qg\to Wq^\prime g$ and 
others) with {\sc Pythia}.  In the
latter case, the showering scale for initial state radiation is the 
same as the factorization scale $Q^2_{ISR} = m^2_W$, while for final state 
radiation it is the same as the invariant mass of the two final state 
partons,  $Q^2_{FSR} = m^2_{jj}$.  Representative color flows are
used to connect the initial state to the final state.   The two
separate calculations are found to be in excellent agreement.  

Jets were found using the clustering code provided in {\sc Pythia}
with a cell size
of $\Delta\eta\times\Delta\phi = 0.1\times 0.1$, a cone radius $R=0.7$ and
a minimum jet $E_T$ of 5 GeV. Cell energies were smeared using a 
calorimeter resolution $\sigma_E$
of $0.5\sqrt{E{\rm (GeV)}}$\cite{resol}. Missing transverse energy $\etmiss$
was estimated by taking the vector sum of the momenta of 
all clusters of energy
found in the calorimeter with $E_T > 5\,$GeV.  In a previous study~\cite{elw}
this was found to give a reasonable representation
of the $\etmiss$ resolution of the D\O\ detector. 

Selected
events were required to have an isolated electron, large missing
energy, and two or more jets. The kinematic selections applied were: 
\begin{itemize}
\item Electron: $E_T > 20\,\gev$; pseudorapidity $|\eta| < 1.0$;
\item Missing energy: $\etmiss > 20\,\gev$;
\item Two or more jets with $E_T > 20\,\gev$;
and $|\eta| < 1.25$, separated from the lepton by at least $\Delta R = 0.7$
\item Leading jet $E_T > 40\,\gev$.
\end{itemize}

Since the technipion was forced to decay into gluons, whose large
color-charge leads to a high probability for final state radiation,
the technipion mass was estimated by:
\[ 
m_{jj(j)} = \left\{ \begin{array}{ll}
\mbox{invariant\ mass}(jet_1,jet_2,jet_3) & \mbox{if $E_T(jet_3)> 15\,\gev$}\\
\mbox{invariant\ mass}(jet_1,jet_2) & \mbox{otherwise}
\end{array}
\right. 
\]
This algorithm seeks to recombine some of the
final state radiation. We have not worked extensively to optimize
our algorithm, since the experimental situation will undoubtedly be more
complicated than we have simulated. 
Our goal is to demonstrate that it is possible
to achieve a better resolution than the naive dijet mass.  
Figure~2 shows
that it gives a significantly improved 
technipion mass resolution compared with
naively taking the invariant mass of the leading two jets, at the cost of
introducing a high-side tail from initial-state radiation.
The peak of the reconstructed mass is shifted downwards from its ``true''
value of 110~GeV by the cumulative effects of final state gluon radiation,
fragmentation (particles emitted outside the cone), and muons and neutrinos
within the jets.  These effects also make the mass resolution much
broader than the calorimetric energy resolution alone would imply.

A similar mass estimation technique might be applicable to the $b\overline b$
invariant mass in such cases as $H \to b\overline b$ searches.
We note, however, that the requirement that there be displaced vertex tags
within each of the candidate $b$ jets already rejects much of the radiative
contamination in the $b\overline b$ case. This is not, of course, possible 
for an object decaying to two gluon jets.  

Requiring that the lepton and jets be central in pseudorapidity exploits
the fact that the signal events will tend
to be produced with larger center-of-mass
scattering angles than the background.  

Figure~3(a) shows the distribution of $m_{jj(j)}$ (as defined above)
jets for signal events (dotted), background (dashed) and their sum (solid)
passing these criteria for a luminosity of $2\,\ifb$.

\begin{figure}
\epsfxsize=15cm
\dofig{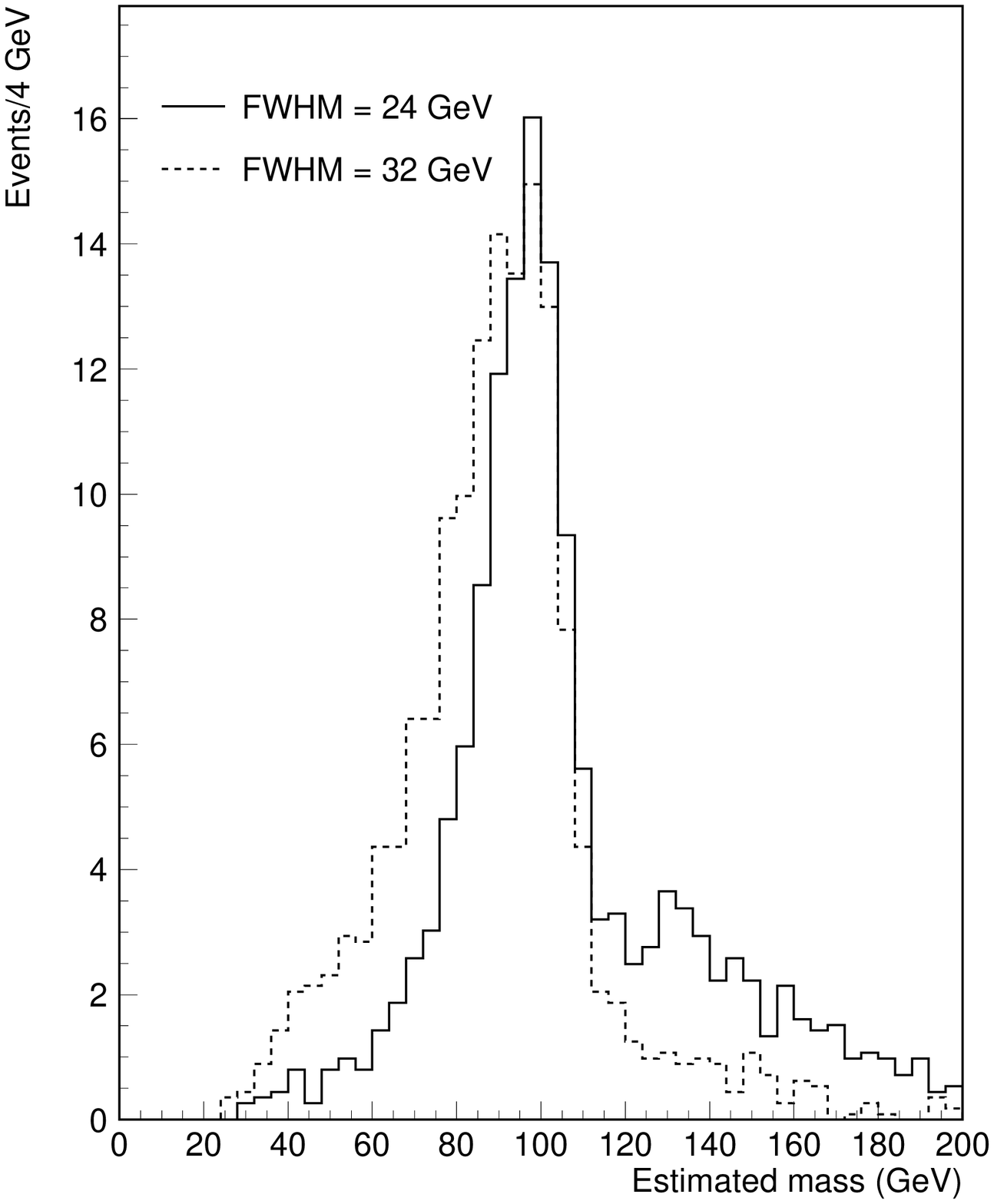}
\caption[]
{Technipion mass resolution in the decay
$\tpiz \to gg$, for signal events passing
the kinematic cuts.  The technipion mass was estimated
firstly as the invariant mass of the two leading jets
(dashed line) and secondly by $m_{jj(j)}$ as defined in
the text (solid line).  The  
full-width at half maximum is reduced from 32 to 24~GeV by
the latter definition.
\label{plot1}}
\end{figure}

\begin{figure}
\epsfxsize=15cm
\dofig{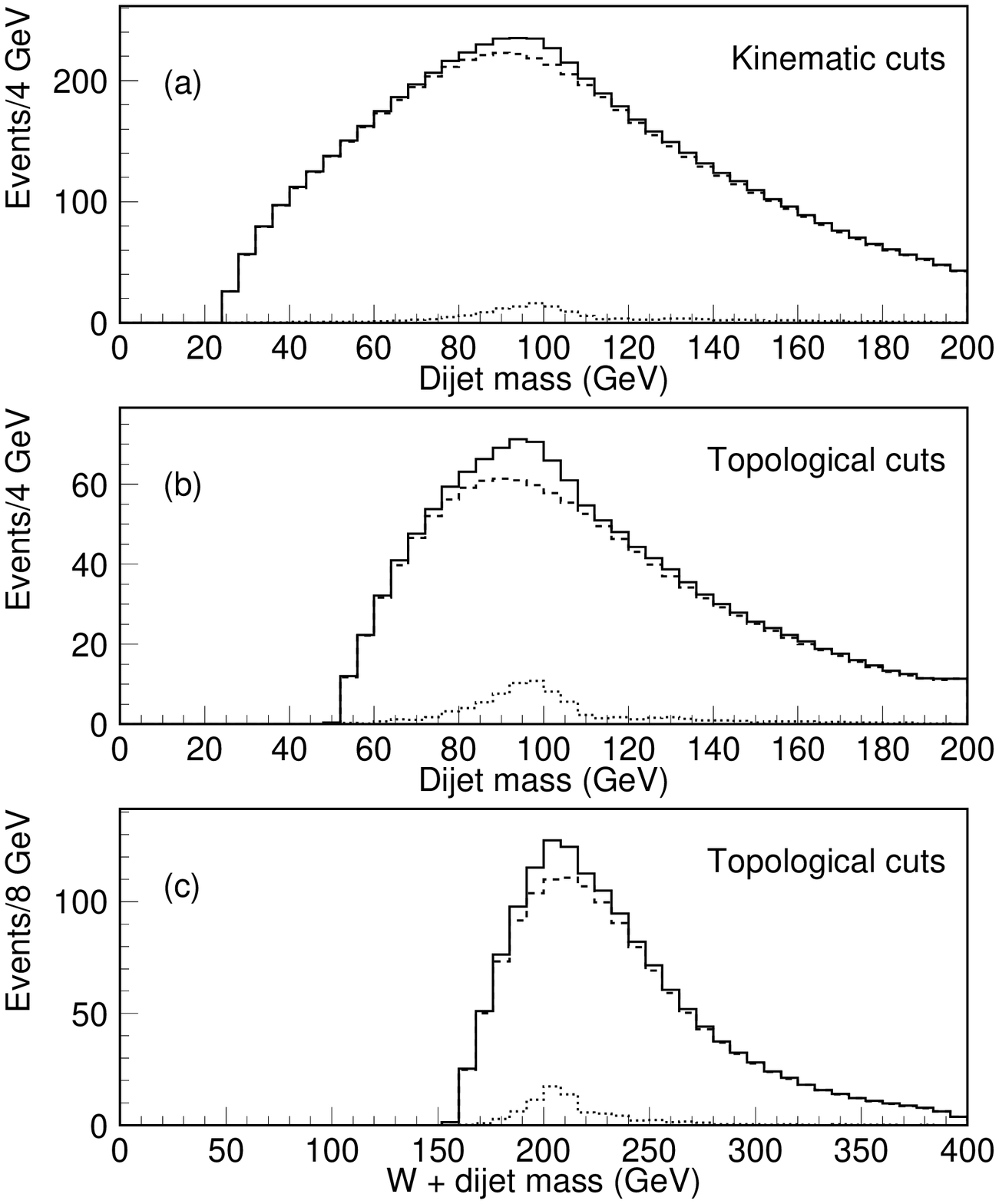}
\caption[]
{Invariant mass distributions for
$\tpiz$ signal (dotted), $Wjj$ background (dashed),
and their sum (solid).  The
vertical scale is events per bin in $2~\ifb$ of
integrated luminosity.
Plot (a) shows the dijet mass $m_{jj(j)}$ with 
kinematic selections only, (b) with the addition 
of topological selections,
and (c) shows the invariant mass of the $W+\tpiz$ system for the same
sample as (b).
\label{plot2}}
\end{figure}

The particular kinematics of $\tro \ra W_L \tpi$ suggests other
cuts that can discriminate signal from the $W+\jets$
background.  The small
$Q$-value for the $\tro$ decay causes the $\tpi$ (and the $W_L$) to have
low longitudinal and transverse momenta, 
and the jets from the technipion decay to be similar in energy and to
have a large
opening azimuthal angle $\dfjj$.
These expectations were borne out by simulated
distributions in these variables.  
Cutting on these variables then helps to suppress the $\wjj$ background to
$\tro \ra W_L \tpi$.

Consequently, we have taken the selected events in Fig.~3(a) and 
applied
additional topological cuts: 
\begin{itemize}
\item $ 35 < p_T^{jj(j)} < 65\,\gev$;
\item $ |p_L^{jj(j)}| < 55\,\gev$;
\item $ (E_T(jet_1) - E_T(jet_2))/(E_T(jet_1) + E_T(jet_2)) < 0.5$;
\item $\dfjj > 90^\circ$.
\end{itemize}
The dijet transverse momentum $p_T^{jj(j)}$ and
longitudinal momentum $p_L^{jj(j)}$
are calculated from the leading three (two) jets if
the third jet has $E_T$ greater (less) than $15\,\gev$, 
just as is the invariant mass $m_{jj(j)}$.
The precise numerical values in these cuts of course depend on the
technirho and technipion masses and their difference.  An experimental
search would need to re-optimize the cut values for each
point in parameter space, as was in fact done in~\cite{cdfsearch}.

The effects of the topological cuts are shown in Fig.~3(b). The 
signal-to-background at $100\,\gev$ is
improved from $\sim 1:12$ to $\sim 1:5$ by these cuts. 
A visible excess is apparent in this distribution, which
corresponds to the $\pi_T^0$ mass.  Also, the signal
distribution peaks in a different mass bin than is expected
for the background.  

Fig.~3(c) shows the corresponding invariant mass distribution
after topological cuts for the $Wjj$ system, which corresponds
roughly to the $\rho_T$ mass.
Here the $W$ four-momentum was reconstructed from the lepton and $\etmiss$,
taking the lower-rapidity solution in each case.  
Unfortunately, while the signal-to-background is also good in the
peak region of this plot, the cuts have resulted in the signal and background
shapes being similar. Unlike the previous figure a deviation from
the expected background shape is not visible.   

To estimate the significance of the signal, we
counted signal $S$ and background $B$ events  
within $\pm 16$ GeV of the peak.
We find $S=85$ and $B=1716$ from Fig. 3(a), yielding
$S/B=.05$ and $S/\sqrt{B}=2.04$.  From Fig. 3(b), we
find $S=54$ and $B=467$, with $S/B=.12$ and 
$S/\sqrt{B}=2.51$.  Including the decays of
the $W$ into muons, and assuming the same efficiency as
for electrons, the significance
of the signal distribution in Fig.~3(b) is increased
to 3.55.  Therefore, if two experiments collect $2~\ifb$
of data in Run II, then a 5 sigma deviation might be observed
in the combined data sets.  Additionally, the nearly degenerate
$\omega_T$ may produce a $\gamma gg$ signal of comparable 
significance.  

In conclusion, we have shown that the low-scale technicolor signature
$\tro \ra W \tpi$ can be discovered 
in Run~II of the Tevatron for production rates as low as a few picobarns,
even if the decay mode $\tpiz \to gg$ dominates.

The research of JW is supported by the Fermi National
Accelerator Laboratory, which is operated by Universities Research
Association, Inc., under Contract~No.~DE--AC02--76CHO3000. 
We thank the Aspen Center for Physics for its hospitality while
this work was carried out.

%
%\noindent Figure 1. 
%Total $WW$, $W\tpi$ and $\tpi\tpi$ cross sections in $\ol p p$
%collisions at $1.8\,\tev$, as a function of $M_{\tro}$ for $M{\tpi} =
%110\,\gev$. The model described above Eq.~\singwidth\ is used with $\sin\chi
%= \third$. The curves are $W^\pm Z^0$ (upper dotted) and $W^+W^-$ (lower
%dotted); $W^\pm \tpiz$ (upper solid), $W^\pm \tpi^\mp$ (lower solid), and
%$Z^0\tpi^\pm$ (long dashed); $\tpi^\pm \tpiz$ (upper short dashed) and
%$\tpip\tpim$ (lower short dashed). EHLQ set 1 distribution functions~\ehlq\
%were used and cross sections were multiplied by a
%$K$-factor of 1.5, as appropriate for Drell-Yan processes.

\end{document}